\documentclass[twocolumn,showpacs,preprintnumbers,amsmath,amssymb]{revtex4}
\usepackage{graphicx}
\usepackage{dcolumn}
\usepackage{bm}

\begin{document}

\preprint{}

\title{Morphological regions and oblique incidence dot
  formation in a model of surface sputtering}   
\author{Emmanuel O. Yewande}
\author{Reiner Kree}
\author{Alexander K. Hartmann}
\affiliation{Institut f\"ur Theoretische Physik,
   Universit\"at G\"ottingen, Friedrich-Hund Platz 1, 37077
   G\"ottingen, Germany 
}

\date{\today}
\begin{abstract}
We study solid surface morphology created by off-normal ion-beam
sputtering with an atomistic, solid-on-solid model
 of sputter erosion. 
With respect to an earlier version of the model, we
 extend this model with the inclusion of lateral 
erosion. Using the 2-dimensional structure factor, we found an
upper bound $\mu\simeq 2$, in the lateral straggle $\mu$,   
for clear ripple formation. Above this upper bound, for
longitudinal straggle $\sigma\gtrsim 1.7$, we found the
possibility of dot formation (without sample rotation). Moreover, a
temporal crossover from a hole topography to ripple
topography with the 
same value of collision cascade parameters  was found. Finally, a scaling
analysis of the roughness, using the consecutive gradient approach,
yields the growth exponents $\beta=0.33$ and $0.67$ for two different
topographic regimes.  
\end{abstract}

\maketitle
\section{\label{sec:intro}INTRODUCTION}
Low energy bombardment of the surface of amorphous materials
\cite{Mayer1994}, semiconductors (amorphized by the sputtering processs)
\cite{Chason1994, Erlebacher1999, Habenicht2002}, and metallic
materials (at low temperature) \cite{Rusponi1998}, by a 
beam of ions at off-normal 
incidence, often lead
to ripple pattern formation. The ripple orientation is perpendicular
to the projection of the ion beam 
direction, onto the surface plane, for small incidence angles, and
parallel to the projection for grazing incidence. However, for a
metallic 
surface with anisotropic diffusion, the orientation is perpendicular
to a crystallographic 
direction (the one favored for diffusion) at small incidence angle
\cite{Rusponi1998}. 

According to the continuum theory \cite{Bradley1988, Makeev2002},
ripples arise, for all angles of incidence $\theta$, from 
the curvature dependence of the sputter yield. 
The basis of the calculation is the 
Sigmund distribution \cite{Sigmund1969} describing the energy
deposited by the incoming ion. This distribution 
results from a study of the collision cascades created by the
penetrating ion. The Sigmund distribution is
parametrized by the depth $a$, the longitudinal straggle $\sigma$,
and lateral straggle $\mu$ of the energy, cf.\ (\ref{eq:Sigmund}).
The wavelength
$\lambda$ obtained within the continuum theory  is given by $\lambda=
2\pi \sqrt{2K/|\nu |}$, where $K$ is the surface diffusivity, and
$\nu$ is a negative  surface tension coefficient, the latter depending
on $a,\mu,\sigma$ and $\theta$. 
Typical wavelengths are of the order of tens of nanometers.
In the absence of non-linearities, the
$\theta$ dependent $\nu$, being 
different along the parallel and perpendicular directions to ion
projection, govern the ripple orientation. However, contrary to the
predictions of the continuum theory 
(ccf. Figs. $12$ and $15$ of \cite{Makeev2002}), 
 no ripples were observed in \cite{Carter1996} for $\theta \lesssim
 40^0$ under Xe ion irradiation of Si.

Other studies have shown that this influence of collision cascading 
on topography evolution extends to
quantum dot formation at $\theta=0$ on non-rotated
substrates\cite{Kahng2001}, and at $\theta>0$ on rotated substrates
\cite{Frost2002}. Also, collision cascade statistics has recently been
found to shift 
\cite{Feix2005} the Sigmund energy distribution away 
from the Gaussian which has been the basis of the theoretical models
so far in use \cite{Bradley1988, Cuerno1995, Hartmann2002}. But till
now the ripple phase boundaries arising from such influence are thought to be
time (i.e. fluence) independent. Furthermore, up till now, dependence of ripple
formation on the lateral straggle $\mu$, though acknowledged, is
often ignored on accounts of isotropic or symmetric cascading 
for simplicity.

Recently a discrete MC model of sputter-erosion was introduced 
\cite{Hartmann2002} which, like the continuum theories,
is based on Sigmund's spatial distribution of the kinetic energy
transfered by an impinging ion. Its very recent application to the
study of ripple motion \cite{Yewande2005} indicates its capture
of the universal features of material-surface modification by ion beam
treatment, and affords us a way of exploring the different
phases in the surface morphology. 

In this paper, we extend this model
with inclusion of lateral erosion. This includes stronger
non-linearities in the sputtering process. Hence,
 the exponential-growth tendency of the
ripple amplitude occuring within the linear theory
is more stabilized \cite{Rost1995, Park1999}. Furthermore, we show that
 the angle $\theta_r$ above which ripple formation occurs depend on the
 longitudinal  straggle $\sigma$ of the ion beam, with higher
 $\sigma$ resulting in lower $\theta_r$.  
Moreover, we find that there are two distinct creation mechanism for ripple:
In one region of the parameter space, there is a
 transition from hole topography at an early time, to ripple
 topography at a later time. For other parameters, the ripples are created
from a simple rough topography at early times.
 Our results also indicate the
 crucial role of the value of $\mu$ for ripple formation. If we chose
$\mu$ larger enough, then 
we find dots/nano-sized islands similar to those observed by 
 \cite{Frost2000, Facsko1999}, and predicted by \cite{Kahng2001,
   Frost2002} without sample rotation. 
Our results suggest that by using
 new projectil/target combinations one might find 
yet unobserved surface topologies. We mention below some
promising projectil/target combinations which might serve as 
guidelines for experimentalists.

In the next section,  we
 describe our simulation model. In the main part 
 we present and discuss our results. We finish with a summary and 
discussion.

\section{\label{sec:algorithm} Numerical simulation methods}
\begin{figure}[!htbp]
\begin{center}
\includegraphics[width=0.4\textwidth]{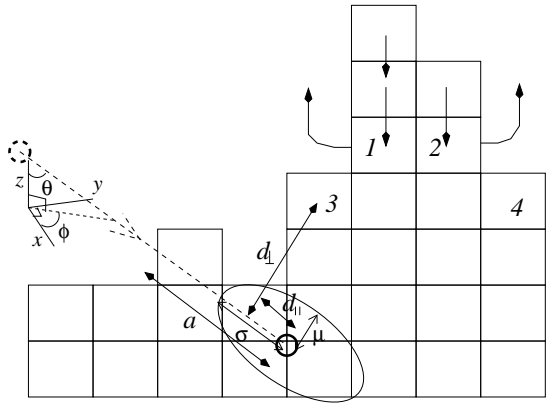} 
\caption{\label{fig:sputter} Sketch showing lateral erosion of
  surface atoms, and collision cascade geometry, as described in the
  text. Erosion of atom at position $1$ or $2$ triggers a relaxation, which 
  ensures there is no overhang; the atom at position $3$ or $4$ is only
  eroded, without surface relaxation.} 
\end{center}
\end{figure}

We study system sizes $L\times L$ with periodic boundary conditions.
The material surface is  defined by a
time-dependent discrete height function $h(x, y, t)$ [$(2+1)D$ solid-on-solid
(SOS) model] which is initially flat, i.e,
$h(x, y, 0)=$constant. Hence, overhangs cannot occur.
We have included surface roughening via sputtering and surface
smoothing via surface diffusion. Each simulation step consist of the
sputtering generated by one ion and a certain number of diffusion
steps. The simulation time is measured in terms of these steps and
corresponds to the fluence in experiments. For details see
Ref.\ \cite{Yewande2005}. Here we only display the main ingredients of
the methods, for the convenience of
the reader, and describe the extensions with respect to the original method.

We simulate the
sputtering process as a combination of ion motion and erosion of
atoms.  The ion source, at some
random position in a 
plane directly above the surface plane, projects the ions along a
trajectory inclined at $\theta$ to the vertical and with 
azimuth $\phi$. After penetrating the solid through a depth $a$,
measured along the ion trajectory, each ion comes to rest and distributes its
kinetic energy $\varepsilon$.
The erosion process is governed by the
generally accepted Sigmund \cite{Sigmund1969} distribution.  
 A surface atom is eroded with a probability 
proportional to the energy deposited there, which is given by 
Sigmunds's formula
\begin{equation}    
E({\bf r})=\frac{\varepsilon}{\sqrt{8\pi^3}\sigma\mu^2}\exp
\biggl\{-\frac{d_\parallel^2}{2\sigma^2}-\frac{d_\perp^2}{2\mu^2}
\biggr\},
\label{eq:Sigmund}
\end{equation}
where $d_\parallel=z+a$ is the distance of the surface atom, from the final
stopping point of the ion, along the ion trajectory. $d_\perp$
is the distance of the surface atom perpendicular to the ion
trajectory. Hence, we use the local coordinate system of the ion with origin
at the point of penetration, and the $z$-axis coinciding with the ion
trajectory \cite{Hartmann2002, Yewande2005}. A
sketch is provided in Fig. \ref{fig:sputter}. 
Note that all parameters depend on the experimental parameters like
materials, ion energy and angle of incidence. For amorphous materials they
can be estimated using the SRIM simulation package \cite{srim}. 
Except where otherwise
stated, $a=6.0$, $\theta=50.0^0$, $\phi=22.0^0$, and $\epsilon$ is
chosen to be $(2\pi)^{3/2}\sigma\mu^2$, while, in contrast to previous studies,
 we vary systematically the values 
of $\mu$ and $\sigma$ mimicing different material combinations.
  
Furthermore, 
we extend the sputtering model by considering lateral erosion of
surface atoms; thus including non-linearity in a manner analogous to
the Kardar-Parisi-Zhang ($KPZ$) non-linearity that arises from lateral
attachment of adatoms 
in surface growth \cite{KPZ1986}. This has the effect of relaxing the surface
\cite{Yewande2006} (expecially as local surface slopes increase, 
since $h(i)\to h(i)-n_i$ where $n_i$ 
is the number of surface atoms eroded in column $i$ at that instant,
due to their being closer to the stopping point of ion than topmost atom
in column (see Fig. \ref{fig:sputter}).

Surface migration is
simulated \cite{Yewande2005, Smilauer1993} as a 
nearest neighbor Monte Carlo 
hopping process, with a site $i$ and nearest neighbor site $j$ chosen
randomly; and a hop $i\to j$ allowed with probability 
\begin{equation}
p_{i\to j}\propto \exp\biggl(- \frac{E}{k_BT}\biggr),
\end{equation}
where $E=E_S+nnE_{NN}+E_{SB}$ is an energy barrier to hopping,
comprising of a substrate 
term $E_S=0.75 eV$, a nearest neighbor term $E_{NN}=0.18 eV$ and a
step barrier term 
$E_{SB}$; $nn$ is the number of in-plane nearest neighbors of the
diffusing atom; $T$ is the 
substrate temperature; and $k_B$ is the Boltzmann's
constant. $E_{SB}=0.15 eV$ if there is no vacant next-nearest neighbors in
the plane below the hopping atom (at site $i$), and there is \emph{at
  least} a vacant next-nearest neighbor in the plane below the hopped
atom (at site $j$); otherwise $E_{SB}=0$. Thus this model discourages
diffusion to a down-step edge but once the approach is made, it does not
discourage a hop down the step (i.e since $E_{SB}=0$ then; see Fig. 1
of \cite{Smilauer1993}). 
A local heating of the surface occurs right after ion
impact, followed 
by rapid cooling; hence surface diffusion is greatly enhanced, due to
higher effective temperatures 
arising from the ion impacts. Consequently, we have used a higher
effective temperature $k_BT=0.1 eV$ in our simulations, 
which was estimated in our previous work \cite{Yewande2005}, based on
a calculation of the spatio-temporal development of the temperature,
arising from the local heating \cite{Marks1997} (see also
\cite{Melngailis1987}).    

\section{\label{sec:results}Results and Discussion}
The profiles shown in the figures are from sizes $L^2=128^2$, and the
bar, on the profiles, indicate the ion beam direction. Distances
are in units of a lattice constant, and time in ions/atom;
except where stated otherwise. 
By varying the collision cascade parameters $\sigma$ and $\mu$, we
explore the full 
topographic features of this sputter-erosion model, for typical ion
energies on the order of $1$KeV; keeping $a$ and
$\theta$ constant. 
 
\begin{figure}[!htbp]
\begin{center}
\includegraphics[width=0.35\textwidth]{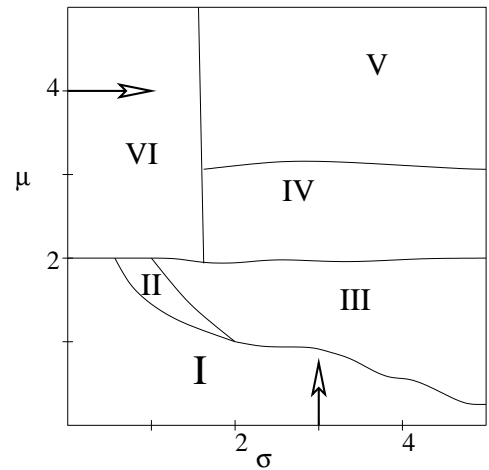} 
\caption{\label{fig:phase} Different topographic regions
  ($\theta=50^0$, $a=6$). Region 
  I: rough surface; II: holes;
  III: clear ripples oriented perpendicular to ion beam direction;
  IV: short ripples (resulting from increased mu); V: dots; VI:
  non-oriented structures. The arrows indicate the directions referred
  to in Fig. \ref{fig:2DSk}.} 
\end{center}
\end{figure}

\begin{figure}[!htbp]
\begin{center}
\includegraphics[width=0.49\textwidth]{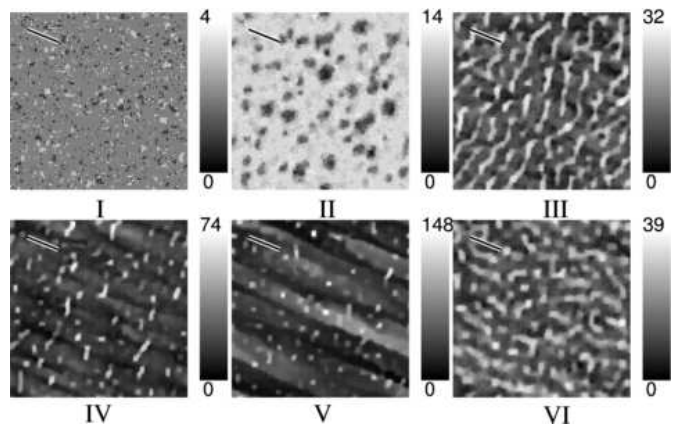} 
\caption{\label{fig:phase_profiles} Profiles for parameters chosen from each
  topographic region in Fig. \ref{fig:phase}; $\theta=50^0$, $a=6$,
  $t=3.0$. (I) $\sigma=1$, $\mu=0.5$; (II) $\sigma=1$, $\mu=1.5$; 
(III) $\sigma=3$, $\mu=1.5$; (IV) $\sigma=4$, $\mu=2.5$; 
(V) $\sigma=5$, $\mu=5$; (VI) $\sigma=0.5$, $\mu=5$} 
\end{center}
\end{figure}

We start with a sketch (Fig. \ref{fig:phase}) of the six topographic
regions that were found,  
coresponding to different 
combinations of $\sigma$ and $\mu$, for $t=3$
ions/surface atom at which almost all the surface 
topographic features are distinct; the corresponding profiles are
shown in Fig \ref{fig:phase_profiles}. Note that the boundaries shown
in this sketch do not represent abrupt transitions from one topography to
another, as we shall see below (in Figs. \ref{fig:q-mu}, \ref{fig:q-sigma}). 
Rather we observe often a smooth crossover from one behavior to the other.
Note that the system is still not in (dynamic) equilibrium \cite{Hartmann2002},
hence the surface morphology still evolves with time and at later time
a diagram of the type of
 Fig. \ref{fig:phase} will look different. We have picked 
a typical time, corresponding to timescales easily accessible in experiments,
which exhibit a rich behavior as a function of the straggling parameters
$\mu$ and $\sigma$.
Also, although the sketch is specifically 
at $\theta=50^0$, similar ``phases'' also occur at other values of $\theta$,
with slight deviations at the boundaries. Examples of a few
experimental parameters (ion 
energy, type of material, type of ion etc.) 
which lead to the specific parameters used here we
obtained
from a SRIM simulation \cite{srim}, are; region V: $1.5-1.7 KeV$
neon(Ne)-ion, or 
$2.25-2.5KeV$ argon(Ar)-ion, sputtering of copper(Cu), or silver(Ag);
$1.2-1.4 KeV$ Ne-ion, or 
$1.7-2.0KeV$ Ar-ion, sputtering of germanium(Ge), or gallium arsenide(GaAs); 
$170-200 eV$ helium(He)-ion sputtering of graphite(C); region IV: 
$650-800 eV$ Ne-ion sputtering of silicon(Si); region III:
$800eV-1.1KeV$ Ar-ion sputtering of silicon(Si); $550-700 eV$ Ne-ion
sputtering of C; with fluences $\sim 10^{14}-\sim 10^{15}$
ions/cm$^2$. Note that 
for most materials and parameter combinations $\sigma \le \mu$, hence the
region VI might be difficult ot access. Also, SRIM
  simulations reveal that very large
  $\sigma$ and $\mu$, i.e, beyond the values considered here, are
  impractical, since they can only occur for a higher $a$. But the
  value of $a$ is itself restricted by the range of ion energies that
  lead to ripple formation.   
We are not aware of a experimental study of the sputtering 
behavior were the parameters are varied systematically in the   
$\mu$-$\sigma$ plane.
Hence, when using these parameters in experiments, 
one might be able to observe the
topographies of Fig. \ref{fig:phase_profiles}, where some of them have not been
observed so far.


\begin{figure}[!htbp]
\begin{center}
\includegraphics[width=0.35\textwidth]{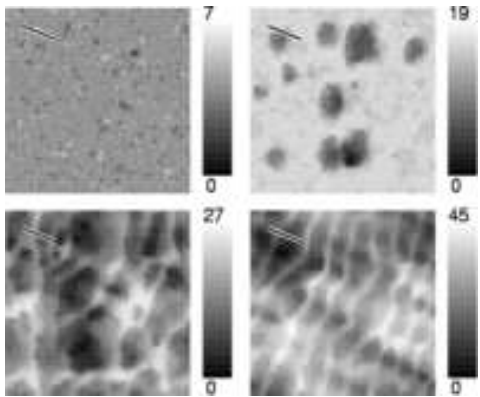} 
\caption{\label{fig:PhaseI-t} Surface profiles of region I,
  $\sigma=1$, $\mu=1$. From top-bottom, left-right, $t=3$, $20$, $40$,
  and $90$.} 
\end{center}
\end{figure}

A brief description of each topographic region in
Fig. \ref{fig:phase_profiles}, including the behavior 
at later times  $t\ge 3$ (atoms/surface atom),
 is as follows:
Region I: rough surface [see Fig. \ref{fig:phase_profiles}(a)] which,
as time increases, evolves to a hole topography. The {\em sizes} of the
holes grow  and finally coalescent 
to a ripple topography at long times (Fig. \ref{fig:PhaseI-t}).

Region II: holes are already prominent in this region [see
Fig. \ref{fig:phase_profiles}(b)]; Here the {\em number}
 of holes in this region increases with time, and again 
 ripples are formed at long times, but at an earlier time than as in
region I (not shown as separate figure).

\begin{figure}[!htbp]
\begin{center}
\includegraphics[width=0.49\textwidth]{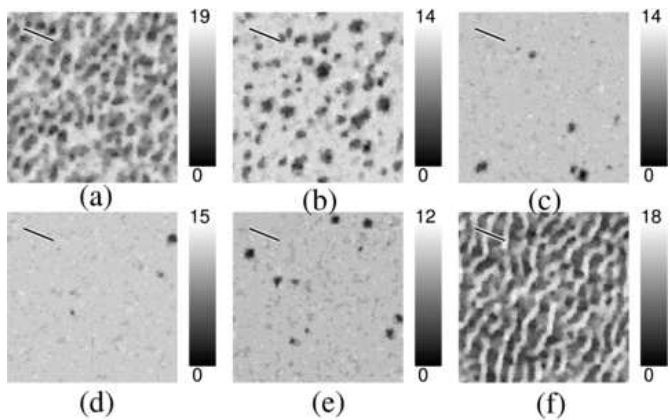} 
\caption{\label{fig:PhaseII} Surface profiles of region II, t=3
  ions/atom ($\sigma=1$, $\mu=1.5$). Top row; $\theta=50^0$, $a=5$
  (a), $a=6$ (b), and $a=7$ (c). Bottom row; $a=6$, $\theta=40^0$ (d),
  $45^0$ (e), and $60^0$ (f).} 
\end{center}
\end{figure}

The number of holes decreases when increasing the sputterin depth $a$
[Fig. \ref{fig:PhaseII} (a)-(c)].
The number of holes also decreases
with decreasing $\theta$ [Fig. \ref{fig:PhaseII} (d)-(f)], while 
ripples are formed already at this early time in this region if
$\theta$ is increased beyond $\theta_r(t=3)\simeq 60^0$.

\begin{figure}[!htbp]
\begin{center}
\includegraphics[width=0.49\textwidth]{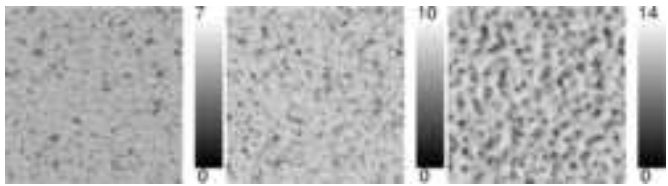}
\caption{\label{fig:PhaseIII} Surface profiles of region III, at
  very early times. From left-right, t=0.1, 0.2 and 0.5
  ions/atom ($\sigma=3$, $\mu=1.5$, $\theta=50^0$, $a=6$).} 
\end{center}
\end{figure}

Region III: the ripple phase \cite{Hartmann2002,
  Yewande2005}. Having observed in regions I and II 
that holes evolve into ripples with time, we studied this
region from the 
very earliest times (t=0-3) but found only very tiny holes, i.e not as
pronounced as in region II, in the course
of ripple formation, see
Fig. \ref{fig:PhaseIII}). $\theta_r\simeq 30^0$ in this region. 
This means, comparing regions I,II and three, there seem to be two different
processes of ripple formation. Ripples can be formed quickly by evolving
directly from a slightly rough surface, or they can be formed slowly via the 
creation of holes, which coalescent to ripples on longer time scales.
Note that in regions I,II, the resulting ripple wavelength is smaller than
the size of the holes generated on smaller time, while in region III the 
ripple wavelength is larger than the tiny holes.

\begin{figure}[!htbp]
\begin{center}
\includegraphics[width=0.49\textwidth]{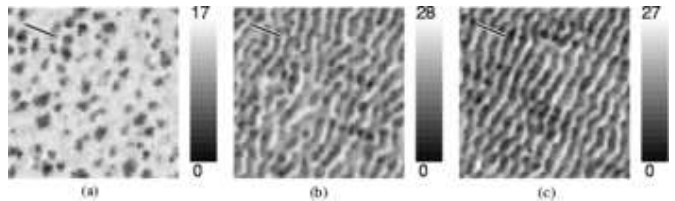} 
\caption{\label{fig:threshold} $\theta=30^0$, $a=6$, $\mu=1.5$;
  $\sigma=2$ (a); $\sigma=3$ (b); and $\sigma=5$ (c).} 
\end{center} 
\end{figure}

We also studied here the case of smaller angles, e.g.\
$\theta=30^\circ$, see Fig. \ref{fig:threshold}. Hence, for lower angles of
incidence, ripple formation is shifted to later times, but  the
ripples tendency again increases when increasing the longitudinal
struggle $\sigma$. This indicates that there might be a lower critical
angle $\theta_c$, below which no ripple formation happens even at long
times. Such an effect has been observed in experiments
\cite{Carter1996} of sputtering 
Xe$^+$ on Si, where below $\theta_C=40^\circ$ no ripples have been found
at finite but long ion fluences. 
For even smaller angles like $\theta=20^\circ$, we indeed do not
observe ripple formation within the time scales (i.e. fluences) we can
reach in our simulations.
Note that a general statement about
the existence of such a critical angle $\theta_c(\mu,\sigma)$ would
require simulations up to very large times for all parameters studied here,
which is beyond the numerical capacities. Hence, we remain with the
statement that our numerical results indicate that such critical angles
indeed exist, without the possibility to determine them precisely.

Region IV: consists of a mixture of dots and short ripples, which
eventually give way to the dot ``phase'' (region V), as $\sigma$ is
increased. Hence, this region seems to ``interpolate'' between regions
III and V.

Region V: consists of dots. These dots are formed on some
ripple-like structures oriented perpendicular to the ion beam
direction, as discussed below within more detail. 
Noting that our model is a solid on
solid model on a square 
lattice, the dots are not unsimilar to the QDs predicted by theory
\cite{Kahng2001, Frost2002}
and observed in experiments \cite{Facsko1999, Gago2001}.

Region VI: consists of non-oriented structures exhibiting a typical
lengthscale, but only a slight orientational preference parallel
to the ion beam. This region, as
mentioned above, 
is probably difficult to access in experiments.

 \begin{figure}[!htbp]
\begin{center}
\includegraphics[width=0.5\textwidth]{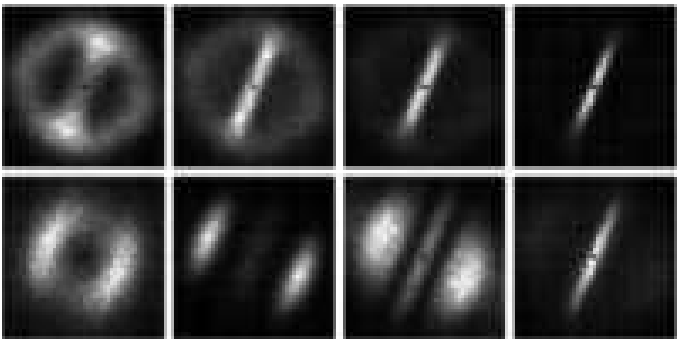} 
\caption{\label{fig:2DSk} $2D$ structure factor (L=128). Top, along $\mu=4$ in
  Fig. \ref{fig:phase}, left-right: $\sigma=0.5$, $1.0$, $1.5$, and
  $3.0$. Bottom, 
  along $\sigma=3$, left-right: $\mu=0.5$, $1.0$, $1.5$, and $3.0$} 
\end{center} 
\end{figure}
\begin{figure}[!htbp]
  \begin{center}
    \includegraphics[width=0.5\textwidth]{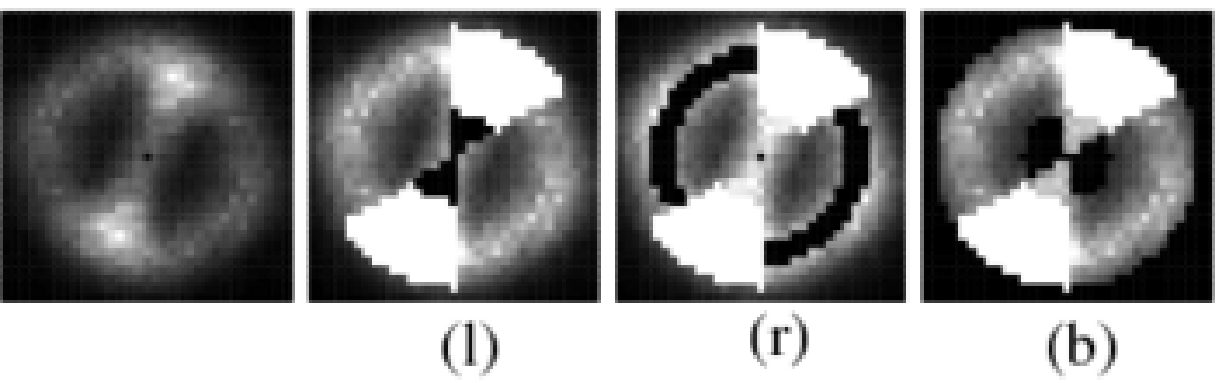}
    \includegraphics[width=0.35\textwidth]{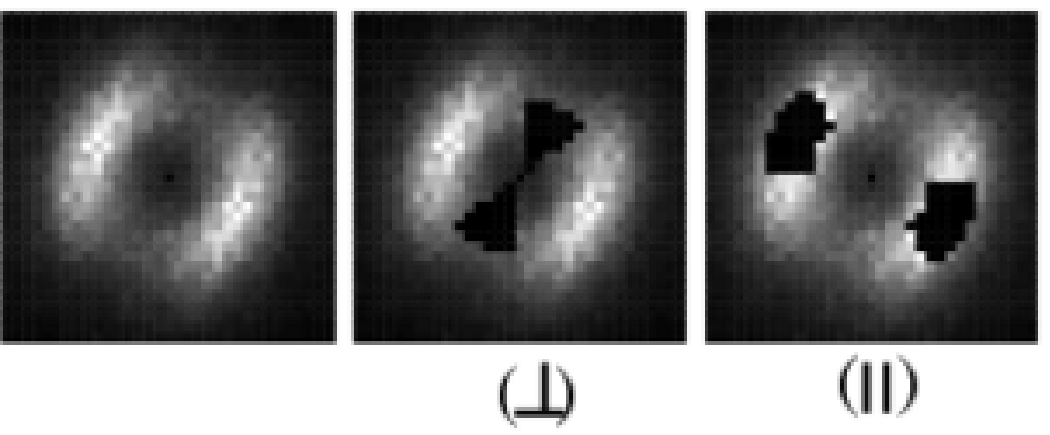}
    \caption{\label{fig:transition_analysis} Separation of the $2D$ S(k)
      (first image top and bottom)
      into regions (in black from second image), for the purpose of
      our analysis. The letters $l$, $r$, $b$,  denote line, ring and
      background, respectively; and the symbols, $\perp$ and
      $\parallel$, denote the perpendicular and parallel directions,
      respectively; as referred to in the text. The white regions are
      excluded from the analysis.}  
  \end{center} 
\end{figure}

A closer look at the dot profiles of region V in Fig. \ref{fig:phase_profiles}
reveals the presence of some underlying large-scale
structures. We now discuss this region and its adjacent regions III
and V with more detail. The underlying structure is clearly seen in a 
$2D$ structure factor, $S({\bf k})=\langle
h({\bf k})h(-{\bf k})\rangle$, 
where h({\bf k}), the fourier transform of the height profile $h({\bf
  r})$, with mean $\langle h\rangle=\sum_{\bf r}h({\bf r})/L^2$, 
at time $t$, is given by
\begin{equation}
\label{eq:FFT}
h({\bf k})=\frac{1}{L}\sum_{\bf r}[h({\bf r})-\langle h\rangle]\exp
(i{\bf k\cdot r}).
\end{equation}
The $2D$ structure factor shown in
Fig.\ \ref{fig:2DSk} has been obtained from an average of 600 independent
runs, for
parameter along $\mu=4$ (top row), and along $\sigma=3.0$ (bottom
row). In these diagrams the structure factor for the
 $k$-vektor $(k_x,k_y)=(0,0)$ is displayed in
the center and the corners represent the values for $(k_x,k_y)=2\pi(\pm
1/8,\pm 1/8)$. 
The case of $\mu=4.0$, when moving from small values of $\sigma$ to
larger values (i.e. left-right in Fig.\ \ref{fig:phase})
is shown in the top row of Fig.\ \ref{fig:2DSk}. For small values of
$\mu$, we are in region IV, where we see a typical wavelength, but
almost no orientation. This translates to a ring visible in the $S(\bf
k)$ plot. Note that there is a slight preference for an orientation
parallel to the ion beam, being visible via two peaks in $S(\bf k)$ at
wave vectors perpendicular to the ion beam. When increasing $\sigma$,
one moves into region V. Here a line perpendicular to the ion beam
emerges in the structure factor rather abruptly 
around $\mu=1$. This line represents the underlying
structures parallel to the ion beam, being visible in in Fig.\
\ref{fig:phase}. Note that the 
  ``dots'' emerge on top of these structures; in the $2D$ structure
factor their signal is too weak to be visible.

 On the other hand, along $\sigma=3.0$, we initially see (bottom line from
bottom-left Fig.\ \ref{fig:phase_profiles}) an orientation, 
spread around ion beam direction (for instance, at
$\mu=1.0$), 
with a more restricted range of $k$  which is typical of the thin wobbly 
ripples. As $\mu$ increases, we
move to the region V as discussed above, but we do not observe an
abrupt change, because for a large range of values of $\mu$, ripples,
dots and the underlying parallel structures coexists (region III).  

To study the crossover or transition from one region to the other in a
quantitative way, it might be more instructive to look at order
parameters which are numbers rahther than the full $2D$ structure factor.
We first
define a quantity {$Q=S^m({\bf k}_{\perp})/S^m({\bf
  k}_\parallel)$}, where $S^m({\bf k}_{\parallel (\perp)})$ is the maximum, for
${\bf k}$ parallel (perpendicular) to ion beam direction. This
quantity detects the change of orientational order.

When moving along $\mu=4$ from region VI into
region V we expect a rather abrupt
change of the behavior from the visual inspection of the $2D$ structure
factor (top row of \ref{fig:2DSk}).
The behavior of $Q$ when changing $\sigma$ for $\mu=4$
is displayed in the inset
 of Fig.\ \ref{fig:q-sigma}. For small values of $\sigma$, i.e. in
 region VI, there is only a slight preference of structures parallel
 to the ion beam (corresponding to ${\bf k}_{\perp}$) which leads to
 small values $Q\approx 2$. When going beyond $\sigma=1$, i.e. when moving
 into region V, the domination of the longitudinal structures
 increases, leading to a growth of $Q$.

We furthermore want to go beyond studying the height of peaks of
 $S(k)$ by examining the presence or absence of a typical
 length scale in the system. This length scale is visible in the $2D$
 structure factor via bright spots (oriented) or a bright ring. For
 this purpose, we
 also look at the relative weights 
\begin{equation}
q_{x/y}=\frac{\int_{A_x} S({\bf k}) d{\bf{}a}}{\int_{A_y} S({\bf k}) d{\bf{}a}},
\end{equation}
of certain areas in the $2D$ structure factor, 
where the integration $d{\bf{}a}$ extends over areas $A_x/A_y$ 
of ${\bf k}$-points. The indices $x,y=l,r,b$, 
refer to  ${\bf k}$ values on the line, 
ring and background (outside line and ring) respectively, as shown in
the upper part 
of Fig. \ref{fig:transition_analysis}. The white regions, in this figure,
are excluded from our analysis, because they contain a superposition
of ${\bf k}$s on both the line, and the ring.
A plot for $q_{r/b}$  is shown in 
Fig. \ref{fig:q-sigma}, other combinations (not shown) yield results
with information not going beyond the study of $Q$.  For $q_{r/b}$
we observe an abrupt change around 
$\sigma=1$ with the ring becoming almost indistinguishable ($q_{r/b}\approx
1$) from the background for $\sigma > 1-1.5$, signaling the
disappearance of a typical length scale around $\sigma=1$. Note that
$q_{r/b}$ stays almost constant for the full region V, in contrast to
the behvior of $Q$.

\begin{figure}[!htbp]
\begin{center}
\includegraphics[width=0.4\textwidth]{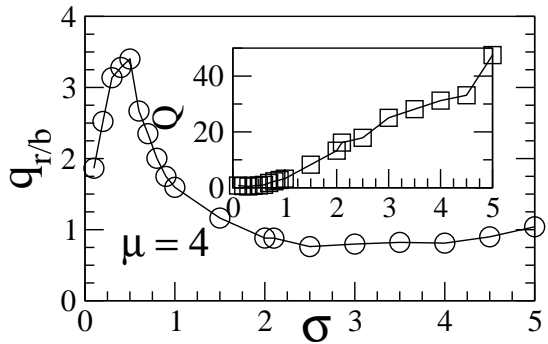} 
\caption{\label{fig:q-sigma} Plot of the quantities $q_{r/b}$ and $Q$ (inset) 
  as a function 
  of $\sigma$, along $\mu=4.0$, as described in the text.} 
\end{center} 
\end{figure}

No we turn to the behavior along the $\sigma=3$ line, i.e. 
when moving from region I, into region III, then IV and finally V. 
The behavior of $Q$ as function of $\mu$ 
(bottom row of Fig.\ \ref{fig:2DSk}) is displayed in the inset
 of Fig.\ \ref{fig:q-mu}. The crossover around $\mu\simeq 2$
from the ripple 
region (II) where $k_\parallel$ wave vectors dominate ($Q<1$)
to the region V
 where $k_\perp$-structures dominate is clearly visible.

The same result is obtained, when again studying not only peaks, but
integrated structure factors over certain selected areas.
Hence, we also study $k$-vectors parallel and perpendicular
to the ion beam ($x,y=\parallel,\perp$), as shown in the lower part of
Fig. \ref{fig:transition_analysis}. Also $q_{\perp/\parallel}$ exhibits
a strong growth for $\mu>2$, i.e. when moving into region III.

Finally, a study of $q_{r/b}$ (not shown) again confirms the
the loss of a certain length scale when moving into region V. This 
is again visible via the disappearance of weight on
 the ring in the $2D$ structure factor with respect to the background
[see again  Fig.\ \ref{fig:2DSk}].
\begin{figure}[!htbp]
\begin{center}
\includegraphics[width=0.4\textwidth]{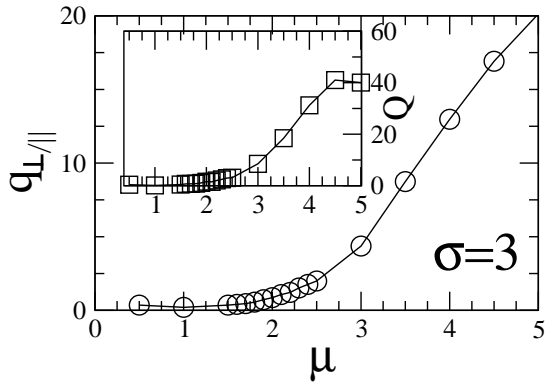} 
\caption{\label{fig:q-mu} Plot of the quantities 
  $q_{\perp/\parallel}$ and $Q$ (inset) 
  as a function 
  of $\mu$, along $\sigma=3.0$, as described in the text.} 
\end{center} 
\end{figure}
 In any case, we have mainly looked at snapshots
at time $t=3$, 
corresponding to typical experimental fluences of $~10^{15}$
ions/cm$^2$ (as already stated above). Hence, the location of the
exact boundaries between the different regions, is not a major
interest here. We only want to demonstrate, that indeed
order-parameter-like functions can be defined. The full information is
anyway contained in the $2D$ structure factors.

In the asymptotic limit, the morphology of sputtered surfaces may be
described by the noisy Kuramoto Sivashinsky equation \cite{Cuerno1995}
 \begin{equation}
\label{eq:KS}
\partial_th=\nu_\parallel \partial^2_x
h+\nu_\perp \partial^2_y h
+\frac{\lambda_\parallel}{2}(\partial_x
h)^2+\frac{\lambda_\perp}{2}(\partial_y h)^2-K\partial^4h+\eta,
\end{equation}
where $\nu_\parallel(\nu_\perp)$, $\lambda_\parallel(\lambda_\perp)$,
 is the effective surface tension coefficient, and 
nonlinear coefficient, respectively,
along a direction parallel (perpendicular) to the projection of the
ion beam direction onto the surface plane (here, the $x$-axis is
parallel to the projection);
 $K$ is the surface diffusion constant; 
and $\eta$ is an 
uncorrelated white noise with zero mean. These coefficients are
explicitly given, in terms of the sputtering parameters, in
ref. \cite{Makeev2002}. Note that erosion of surface material tends to
maximize the exposed area, so-called negative surface tension, hence,
for the sputtering phenomena, this instability constrains the
coefficients $\nu_\parallel$ and $\nu_\perp$ to be negative.

\begin{figure}[!htbp]
  \begin{center}
    \includegraphics[width=0.4\textwidth]
    {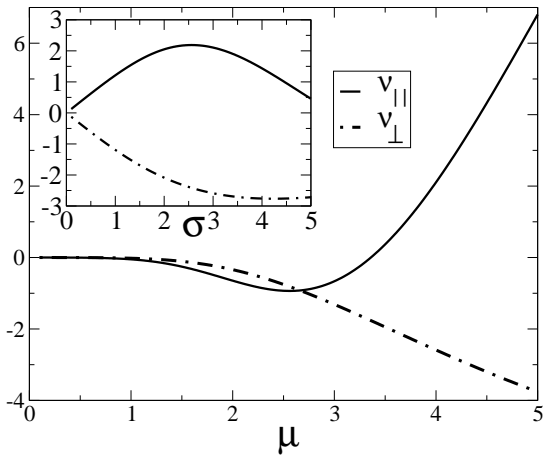}   
    \caption{\label{fig:instability_coeff} The surface tension
      coefficients of Eq. \ref{eq:KS}, along $\sigma=3.0$, for our
      simulation parameters. $\theta=50^\circ$, $a=6.0$}  
  \end{center}
\end{figure}

At early times, the local slopes are small enough in most of the
regions for us to ignore the
nonlinearities; we are therefore left with a noisy Bradley-Harper
equation \cite{Bradley1988}. A plot of the coefficients $\nu$ for our
parameter range, along $\sigma=3$, is shown in 
Fig. \ref{fig:instability_coeff}; $|\nu_\parallel|$ and $|\nu_\perp|$
are nonzero always which implies the presence of two
lengthscales. Hence, according to the linear continuum theory, 
ripples parallel and perpendicular to the ion beam direction are
always present in the system; the one observed is the one for which
$|\nu|$ is highest; i.e, with the highest amplitude-growth rate
$R=-(\nu_\parallel k_\parallel^2+\nu_\perp
k_\perp^2)-K(k_\parallel^2+k_\perp^2)^2$ \cite{Bradley1988}. 

We now try to understand roughly the behvior along the above discussed
$\sigma=3$ and $\mu=4$ lines within the linear theory. For a full
understanding one would have to consider also the nonlinear terms,
where the full dependance on the parameters is yet not available to
the authors.
In Fig. \ref{fig:instability_coeff} the
values of $\nu_\parallel$ and $\nu_\perp$ are shown as function of
$\mu$ (main plot) and $\sigma$ (inset) along these lines.
For the case $\mu=4$ (see inset), we observe that $\nu_\parallel>0$ for
all values of $\sigma$, hence the preferential orientation is always
parallel to the ion beam, as observed. To understand the crossover
from region VI to region V, one has probably to consider the nonliear
terms. 

For the $\sigma=3$ line, we observe $|\nu_\perp|\approx |\nu_\parallel|$ for $\mu
<1$, which is compatible with the behavior in region I, where no
preferential orientation is observed.
For $1 \lesssim \mu \lesssim 2.5$,
$|\nu_\parallel|>|\nu_\perp|$ which implies the dominance of ripples
with $k_\parallel$ (i.e, with wavelength $\lambda=2\pi
\sqrt{2K/|\nu_\parallel |}$), as we have seen in region $III$ and partly
in the crossover region IV. In region
IV ripples are still present, but dominate less.
For $\mu$ 
above this range, the structures with $k_\perp$ dominate. However, the
positive value of $\nu_\parallel$ for higher $\mu$ is contrary to our results,
since it implies a preferential smoothening along the parallel
direction. We have not observed such smoothening,  here probably
nonlinear effects 
are more important. If we consider
the maximum of $|\nu_\parallel|$ and $|\nu_\perp|$ for the region
where $\nu_\parallel<0$ and/or $\nu_perp<0$, then
we see that this maximum is obtained in region V for large $\mu$.
Indeed we observe that region V is rougher
than e.g.\ region III, see below.

\begin{figure}[!htbp]
  \begin{center}
    \includegraphics[width=0.4\textwidth]
    {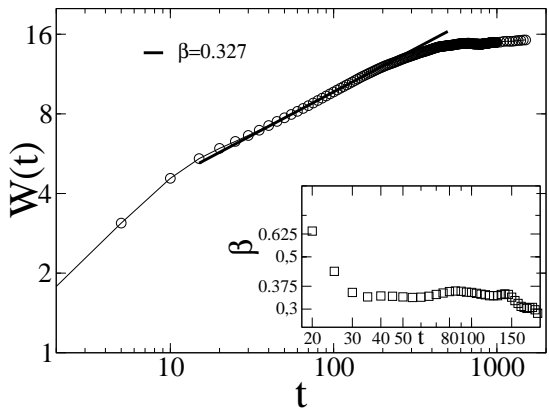}   
    \caption{\label{fig:roughness_beta} Surface width $W$ versus
      time, $\sigma=3$, $\mu=1.5$ (region III), $\theta=50^0$. In the inset is
      a plot of the consecutive $\beta$s for $t=20-200$, where the lower and
      upper cutoffs are seen to occur at $t=35$ and $150$ respectively.} 
  \end{center}
\end{figure}
\begin{figure}[!htbp]
  \begin{center}
    \includegraphics[width=0.4\textwidth]
    {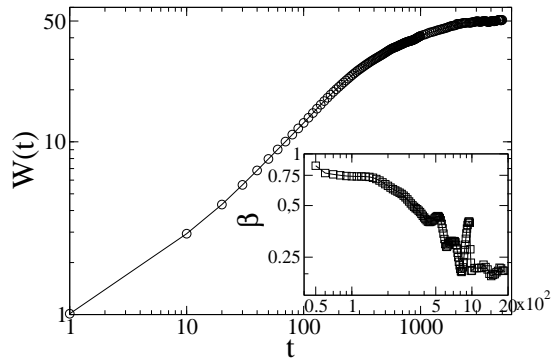} 
    \caption{\label{fig:roughness_beta2} Surface width $W$ versus
      time, $\sigma=5$, $\mu=5$ (region V), $\theta=50^0$. In the inset,
      time is in multiples of $10^2$; there are two scaling regions
      $t=40-140$ and $t=1000-1800$, which 
      yields $\beta=0.665$ and $0.137$ respectively.} 
  \end{center}
\end{figure}

But, before we proceed, we should recall that regions I and II
eventually evolve 
to a similar ripple topography as in region III; region IV 
is an intermediate stage
and VI  is hardly experimentally accessible. Therefore, 
we are interested in two different scaling
regions, i.e, arising from ripple-yielding, as well as dot yielding
parameters. We use the finite-size scaling behavior
 of the surface width/roughness, 
\begin{equation}
\label{eq:fss}
W(L, t)= L^\alpha f(\frac{t}{L^z}) ,
\end{equation}
where $\alpha$ and $\beta$ are the roughness exponent and growth
exponent, respectively; and $W=\sum_{\bf r}(h({\bf r})-\langle h
\rangle)^2/L^2$, i.e, $W$ is the root-mean-squared fluctuations in the
surface height. (\ref{eq:fss})
 defines two scaling regions, separated at the saturation time
$t_s(L)\propto L^z$. $z=\alpha/\beta$ is the dynamic exponent.
In the first region
($t<<L^z$) $W\propto t^\beta$, 
independent 
of $L$; and in the second region ($t>>L^z$), $W\propto L^\alpha$,
constant in time \cite{Barabasi1995}.
{A scaling argument in \cite{Park1999} reveals that $t_s$
  increases with the diffusion coefficient, and with decreasing strength
  of non-linearities \cite{DS1991}.} 

In the following roughness analysis, we rescale the time unit such that
 $L^2$ particles are eroded in unit time,
which is analogous to the measure of time in simulations of epitaxial
growth on vicinal surfaces.

We obtained our scaling exponents from an average of
$600$ independent runs. Fig. \ref{fig:roughness_beta} is a 
plot of the surface width $W(t)$ for $L=128$, we determine the actual 
scaling region of $\beta$ from the
consecutive slopes \cite{Barabasi1995} of W(t) for $t<L^z$; shown in
the inset of Figs. \ref{fig:roughness_beta} and
\ref{fig:roughness_beta2} where a 
 fit to this region gives $\beta=0.327\pm 0.001$ and $\beta=0.665 \pm
 0.003$ (from the scaling for times around $t=100$)
for the topographic regions (see Fig. \ref{fig:phase}) III and
 V respectively. 

Our exponents are
quite different from the scaling of the KPZ equation \cite{KPZ1986}  in (2+1)
dimensions \cite{Amar1990,Barabasi1995}, which shows that for these parameter
choices, either one or 
both of the negative surface tension coefficients, { in the
  noisy $KS$ equation \cite{Cuerno1995}}, do not renormalize to
positive values. { That is, if indeed, this atomistic
  model which is known to accurately describe the morphology of
  sputtered surfaces at early and long times 
  \cite{Hartmann2002,Yewande2005,Brown2005}, can be represented by
  the noisy $KS$ equation, in the asymptotic limit, as expected for
  the sputtering phenomenon.}   
\section{Conclusions}
We have extended the study of surface topography induced
by ion bombardment beyond the circular-symmetric collision cascade case
$\sigma=\mu=1$; using a 
discrete Monte Carlo model. We found an upper bound $\mu\simeq 2$ for
clear ripple formation.
We observed a crossover from hole topography to ripple topography,
for same collision cascade parameters. 
Above the upper bound we found a crossover to a dot
topography; our results indicate the non-trivial influence  of 
the value of $\mu$
for the surface topography. Moreover, we found the possibility of dot
formation for $\mu > 2$ and $\sigma\gtrsim 1.7$, without sample rotation.

Also, we found the possibility of 
different incidence $\theta_r$, around which ripple formation is
possible, as in 
the experiment; our results indicating that high value of the 
longitudinal
straggling result in lower value of 
$\theta_r$. Finally, we found that the collision
cascade parameters affect the growth exponent.  

Our results indicate that using new target/projectil combinations
different in experiments might lead to surface topographies yet not observed. 
Here a systematic study covering the full parameter space 
(at least in the range ($\sigma \le \mu$) would be very interesting, which has
not been done so far to our knowledge.
Furthermore, it would be nice to investigate the small-time behavior, i.e.\
the ripple formation process, experimentally. In this way one could verify 
whether the two   different creation 
mechanism we have observed, hole-coalescence and
creation from a rough surface, can be observed in experimental systems.

\begin{acknowledgments} 
The large scale numerical simulations were performed on the
workstation clusters of the i nstitute.  
This work was funded by the DFG
({\em Deutsche Forschungsgemeinschaft}) 
within the SFB ({\em Sonderforschungbereich}) 602:{\em Complex Structures in
Condensed Matter from Atomic to Mesoscopic Scales} and by
the {\em VolkswagenStiftung} (Germany) within the program
``Nachwuchsgruppen an Universit\"aten''.
\end{acknowledgments}


\begin{thebibliography}{99}
\bibitem{Mayer1994} T. M. Mayer, E. Chason, and A. J. Howard,
  J. Appl. Phys. {\bf 76}, 1633 (1994).

\bibitem{Chason1994} E. Chason, T. M. Mayer, B. K. Kellerman,
  D. T. McIlroy, and A. J. Howard, Phys. Rev. Lett. {\bf 72}, 3040 (1994).

\bibitem{Erlebacher1999} J. Erlebacher, M. J. Aziz, E. Chason,
  M. B. Sinclair, and 
  J. A. Floro, Phys. Rev. Lett {\bf 82}, 2330 (1999).

\bibitem{Habenicht2002} S. Habenicht, K. P. Lieb, J. Koch, and A. D. Wieck, 
Phys. Rev. B {\bf 65}, 115327 (2002).

\bibitem{Rusponi1998} S. Rusponi, G. Costantini, C. Boragno, and U. Valbusa,
  Phys. Rev. Lett. {\bf 81}, 2735 (1998). 

\bibitem{Gago2001} R. Gago, L. V\'azquez, R. Cuerno, M. Varela,
  C. Ballesteros, and J. M. Albella, Appl. Phys. Lett. {\bf 78}, 3316
  (2001).

\bibitem{Bradley1988} 
R. M. Bradley and J. M. E. Harper, J. Vac. Sci. Technol. A {\bf
  6}, 2390, (1988); and references therein.

\bibitem{Makeev2002} 
M. Makeev, R. Cuerno and A. -L. Barab\'asi, Nuc. Instr. and
  Meth. in Phys. Res. B {\bf 197}, 185 (2002).

\bibitem{Carter1996} G. Carter, and V. Vishnyakov, Phys. Rev. B {\bf 54}, 17647
  (1996).

\bibitem{Kahng2001} B. Kahng, H. Jeong, and A.-L. Barab\'asi,
  Appl. Phys. Lett. {\bf 78}, 805 (2001).

\bibitem{Frost2002} F. Frost, Appl. Phys. A {\bf 74}, 131 (2002).

\bibitem{Feix2005} M. Feix, A. Hartmann, R. Kree, J. Mu$\hat{n}$oz-Garc\'ia,
  and R. Cuerno, Phys. Rev. B 71, 125407 (2005).

\bibitem{Sigmund1969}  P. Sigmund, Phys. Rev. {\bf 184}, 383 (1969).

\bibitem{Cuerno1995} R. Cuerno and A. -L. Barab\'asi,
  Phys. Rev. Lett. {\bf 74}, 4746 (1995).

\bibitem{Hartmann2002} 
  A. K. Hartmann, R. Kree, U. Geyer, and M. K\"olbel, Phys. Rev. B
  {\bf 65}, 193403 (2002).

\bibitem{Yewande2005} E. O. Yewande, A. K. Hartmann, and R. Kree,
  Phys. Rev. B {\bf 71}, 195405 (2005).

\bibitem{Rost1995} M. Rost and J. Krug, Phys. Rev. Lett. {\bf 75},
  3894 (1995).

\bibitem{Park1999} S. Park, B. Kahng, H. Jeong, and A.-L. Barab\'asi,
  Phys. Rev. Lett. {\bf 83}, 3486 (1999).

\bibitem{Frost2000} F. Frost, A. Schindler, and F. Bigl,
  Phys. Rev. Lett {\bf 85}, 4116 (2000).

\bibitem{Facsko1999} S. Facsko, T. Dekorsy, C. Koerdt, C. Trappe, H. Kurz,
  A. Vogt, and H. L. Hartnagel, Science {\bf 285}, 1551 (1999).

\bibitem{srim} J.F. Ziegeler, J.P. Biersack and K. Littmark, 
{\em The Stopping and Range of Ions in Matter}, (Pergamon, New York
1985); see also
  \verb!http://www.srim.org/!.

\bibitem{KPZ1986} M. Kardar, G. Parisi, and Y.-C. Zhang,
Phys. Rev. Lett. {\bf 56}, 889 (1986).

\bibitem{Yewande2006} E. O. Yewande, unpublished.

\bibitem{Smilauer1993} P. $\breve{S}$milauer, M. R. Wilby, and D. D. Vvedensky,
  Phys. Rev. B {\bf 47}, 4119 (1993).

\bibitem{Marks1997} N.A. Marks, Phys. Rev. B {\bf 56}, 2441 (1997)

\bibitem{Melngailis1987} J. Melngailis, J. Vac. Sci. Technol. B {\bf
    5}, 469 (1987).

\bibitem{Barabasi1995} A. -L. Barab\'asi and H. E. Stanley,
  \emph{Fractal Concepts  
in Surface Growth} (Cambridge University Press, Cambridge, 1995).

\bibitem{DS1991} See also: Z.-W. Lai and S. Das Sarma,
  Phys. Rev. Lett. {\bf 66}, 2348 (1991); where another non-linearity,
  distinct from the $KPZ$ non-linearity was shown to affect the
  surface topography, in conserved $MBE$
, in a similar manner.

\bibitem{Amar1990} J. G. Amar and F. Family, Phys. Rev. A {\bf 41}, 3399 (1990).

\bibitem{Brown2005} A.-D. Brown, J. Erlebacher, W.-L. Chan, and
E. Chason, Phys. Rev. Lett. {\bf 95}, 056101 (2005).
\end{thebibliography}
\end{document}